# Transduction of an immortalized olfactory ensheathing glia cell line with the green fluorescent protein (GFP) gene: evaluation of its neuroregenerative capacity as a proof of concept.

Plaza N*, Simón D*, Sierra J, Moreno-Flores MT[&]

*   These authors have contributed equally to this work.

Neuroregeneration Group, Biotechnology Department. Universidad Francisco de Vitoria, Ctra. Pozuelo-Majadahonda, Km. 1.900, Pozuelo de Alarcón, 28223 Madrid, Spain.

[&]Corresponding author / present address:
MT Moreno-Flores
Anatomy, Histology and Neuroscience Department
Faculty of Medicine
Universidad Autónoma de Madrid
C/ Arzobispo Morcillo, 2, 28029 Madrid, Spain.
e-mail: mteresa.moreno@uam.es / mt.moreno.prof@ufv.es

**Abbreviations:**
CNS: central nervous system; E-GFP: Enhanced green fluorescent protein; FACS: fluorescence-activated cell sorting; FCS: Foetal calf serum; GFP: green fluorescent protein; OEG: olfactory ensheathing glia; ihOEG: immortalized human olfactory ensheathing glia; LargeT-SV40: Large T antigen from SV40 virus; MAP: Microbutule-associated proteins; PFA: paraformaldehyde; PLL: poly-L-lysine; RGN: retinal ganglion neuron; S.E.M.: Standard error of mean; TERT: telomerase catalytic subunit.

**Abstract**

Olfactory ensheathing glia (OEG) cells are known to foster axonal regeneration of central nervous system (CNS) neurons. Several lines of reversibly immortalized human OEG (ihOEG) have been previously established that enabled to develop models for their validation *in vitro* and *in vivo*. In this work, a constitutively GFP-expressing ihOEG cell line was obtained, and named Ts14-GFP. Ts14-GFP neuroregenerative ability was similar to that found for the parental line Ts14 and it can be assayed using *in vivo* transplantation experimental paradigms, after spinal cord or optic nerve damage. Additionally, we have engineered a low-regenerative ihOEG line, hTL2, using lentiviral transduction of the large T antigen from SV40 virus, denominated from now on Ts12. Ts12 can be used as a low regeneration control in these experiments.

**Key words:** Immortalized human olfactory ensheathing glia (ihOEG); retinal ganglion neuron (RGN); axonal regeneration; green fluorescent protein (GFP); proof of concept.

**Introduction**

From Cajal's classical studies, it is known that in contrast to neurons from the peripheral nervous system (PNS), central nervous system (CNS) neurons regenerate poorly [1]. Several strategies have been used to promote CNS regeneration [2], but during the last years the use of olfactory ensheathing glía (OEG) for CNS regeneration has received much attention due to their special properties [2-7]. These cells have been described to derive either from the olfactory placode [8] or, in contrast, the neural crest [9]. Since throughout life OEG usually surrounds olfactory axons growing into the adult mammalian CNS, it was reasonable to hypothesise that they might facilitate axonal regeneration [2-4,6,10-13]. In fact, several studies have confirmed their neuroregenerative and neuroprotective capacities in different models of CNS injury [14-19]. This OEG reparative capacity is due to a combination of several factors [2,3], including inflammation/angiogenesis control, astrocytes interaction [2,20-22], glial scar decrease [2,20], extracellular matrix proteases production [23], secretion of neurotrophic [24-26] and axonal growing factors [27], and ability to promote myelination [28].

Given the technical limitations to expand primary OEG, maintaining the regenerative properties for transplantation [3], we have previously described the establishment of several immortalized OEG clonal lines (iOEG) derived from primary cultures prepared from rat olfactory bulb [29]. These lines provide an unlimited supply of homogeneous OEG cells and maintain the regenerative capacity of the original cultures both i) *in vitro*, in a model of adult axonal regeneration in cocultures of OEG lines with adult RGNs [23,29,30]; and ii) *in vivo* in a model of spinal cord injury and transplantation [31]. We extended these studies to human OEG (hOEG), describing the preparation of primary cultures from olfactory bulbs obtained in autopsies. These primary cells were reversibly immortalized using the genes of telomerase catalytic subunit (TERT) and Bmi-1, obtaining several ihOEG cell lines that conserved their neuroregenerative properties (ihOEG hTLs) [32,33]. hTL4 cells were additionally modified with the SV40 virus large T antigen [34], originating in this way a Ts14 OEG cell line, with rapid growth capacity. This line provided an excellent tool for the study of OEG properties *in vitro* [34].

Previously, we also demonstrated that hTL4 cells engineered to constitutively express green fluorescent protein (GFP) can be tracked several weeks after xenotransplantation in the injured spinal cord of nude rats [32]. In the future, we aim to follow Ts14 cells after transplantation in the injured CNS (i.e. spinal cord or optic nerve). We describe here: i) the additional modification of Ts14 line, engineering it to constitutively express GFP; and ii) the generation of a modification of the previously described [32] low-regenerative ihOEG line hTL2, using lentiviral transduction of the SV40 virus large T antigen, thus originating the Ts12 ihOEG line. Using our *in vitro* coculture model as a proof of concept, we have determined preservation of i) neuroregenerative capacity of Ts14-GFP line in comparison with its parental line Ts14, and ii) the low-regenerative capacity of Ts12 line.

## 2. Materials and methods

*2.1. Animals*

All animal experimentation was carried out in animal facilities of UFV complying with the Spanish Royal Decree 223/1988, which follows the European Council Directive 86/609/EEC (1986), and approved by national and institutional bioethics committees. Animals were maintained on a 12-hour light/12-hour dark cycle in a day, and were supplied with regular food and water *ad libitum*.

*2.2. Cell cultures*

ihOEG Cell line Ts14 was maintained as previously described, using ME10 [29,30], composed by DMEM-F12 medium supplemented with 10% foetal calf serum (FCS) Hyclone (ThermoScientific), 2mM glutamine (Thermoscientific), 20 µg/mL pituitary extract (Gibco, Life Technologies), 2 µM forskolin (Sigma) and antibiotics (Primocin 100 µg/mL and Plasmocin 25 µg/mL, Invivogen) at 37ºC in 5% $CO_2$.

*2.3. Immunostaining*

Cells were plated in coverslips and after 24-48 hours were fixed with 4% paraformaldehyde (PFA) in phosphate buffered saline (PBS). After several washes with PBS, immunocytochemistry was performed. Briefly, cells were blocked with 0,1% Triton X-100/1% FCS in PBS (PBS-TS). Primary antibodies were prepared in this buffer as follows: mouse monoclonal antibodies against S100β (1:500, SIGMA), Neuroligin-3 (1:1000, Synaptic Systems), SV40 Large T antigen (1:250, Pharmingen) and GFP (1:1000, Cell Signalling); and rabbit polyclonal antibody against GFAP (glial fibrillary acidic protein, 1:1000; Chemicon). After several washes with PBS, cells were incubated with the corresponding fluorescent secondary antibodies prepared in PBS-TS (conjugated with either alexa 488 or alexa 594 fluorophores). Finally, coverslips were washed and mounted with Fluoromount (Southern Biotech, Birmingham, AL).

In some cases, fluorescent nuclei staining with DAPI (4',6-Diamidino-2-phenylindole) was performed after incubation with secondary antibodies. After washing, cells were incubated for 10 minutes in dark with DAPI (10 µg mL$^{-1}$ in PBS/0,1% FCS/0,01% Triton X-100). Then, coverslips were washed and mounted with Fluoromount.

*2.4. Lentiviral vectors packaging using HEK-293T cells*

The packaging plasmid pCMVdR8.74 [35] and the vesicular stomatitis virus G envelope protein plasmid pMD2G (Addgene plasmid 12259) were kindly supplied by Dr. Filip Lim. Production and purification of pRRLSIN.cPPT.PGK-GFP.WPRE vector (Addgene plasmid 12252) was performed using Quiagen columns and later plasmid preparations were assayed with restriction endonucleases and electrophoresis in agarose gel (data not shown).

Lentivectors encoding E-GFP and large T antigen from SV40 (T-SV40) were produced by transient co-transfection of 5 µg of the pRRLSIN.cPPT.PGK-GFP.WPRE vector or pLOX-Ttag-iresTK vector, which express E-GFP and SV40 large T antigen, respectively; 5 µg of the packaging plasmid pCMVdR8.74 and 2 µg of the plasmid pMD2G in 10-cm plates of sub-confluent HEK 293T cells, using Lipofectamine Plus reagent following instructions of the supplier (Invitrogen). After 48 hours, supernatant with the infectious particles was recovered, aliquoted and frozen at -80ºC. Lentivectors were titered on target cells (hOEG) with serial dilutions of the vector supernatants, and the number of transduced cells was determined 48 hours post-infection by flow cytometry.

*2.5. Lentiviral Infection and Ts14-GFP and Ts12 selection*

Ts14-GFP and Ts12 cell lines were generated by infecting Ts14 and hTL2 lines with E-GFP and LargeT-SV40 encoding lentivirus, respectively. A multiplicity of infection (MOI) of 10 infectious units/cell was used, and incubation was performed during 4-6 hours in DMEM-10% FCS (M10), without tissue extracts.

Ts14 cells were washed in M10 and maintained for 48 hours, for transgene expression. Ts14 cells exhibiting high E-GFP expression were selected by flow cytometry (FACS: *Fluorescence-Activated Cells Sorting; FACSVantage SE*), and this line was named as Ts14-GFP. In the case of infected hTL2 cells, they were washed in M10, maintained in M10 for two weeks after the infection, and then changed to complete medium ME10 (with tissue extracts), until a population of rapidly growing cells was selected. This line was named as Ts12.

Both cell lines, Ts14-GFP and Ts12 were tested to assess their neuroregenerative capacity *in vitro*, using Ts14 as a positive control (coculture model, proof of concept).

*2.6. Axonal Regeneration Assay in vitro: coculture of RGN with ihOEG (Ts14, Ts14-GFP and Ts12)*

Adult RGN regenerate their axons very poorly in culture on poly-L-Lysine (PLL), but plated onto OEG monolayers are able to outgrowth many neurites (dendrites and long axons). Cocultures of adult rat RGN (p60 animals) with monolayers of ihOEG, Ts14, Ts14-GFP and Ts12, were performed as previously described [23,29,36]. Briefly, retinal tissue was extracted from adult 2 month-old rats and digested with papain (20 U/ml papain; Worthington, Lakewood, NJ) in the presence of 50 µM of the NMDA receptor inhibitor, D,L-2-amino-5-phosphonovaleric acid (Sigma). The mixed retinal cell suspension was then plated on either 10 µg/ml PLL-treated coverslips or onihOEG: Ts14, Ts14-GFP and Ts12 monolayers. Cultures were maintained at 37°C with 5% $CO_2$ for 96 hr. in serum-free Neurobasal medium (Invitrogen, Carlsbad, CA, USA) supplemented with B-27 (Invitrogen) and 25 mM KCl (NB-B27), before they were fixed with 4% PFA in PBS. Then, immunocytochemistry for axonal (SMI31 epitope in MAP1B and neurofilament) and somatodendritic (MAP2A/B, 514 antibody) markers, was performed.

Axonal regeneration was quantified as percentage of neurons with axon (SMI31 positive neurite) respect to total population of RGNs (identified with MAP2A/B, 514 positive immunostaining of neuronal body and dendrites). Additionally, mean axonal length was determined using the application NeuronJ of the software ImageJ (Wayne Rasband, National Institutes of Health, USA). Axonal regeneration index was calculated as mean axonal length (µm)/neuron. Neuron adherence/survival values were estimated by counting the number of neurons per field (magnification x400). At least 200 neurons or 20 fields were quantified by taking a picture in a CCD monochrome and colour (Spot ST, Slider), coupled to a microscope Axiovert200 (Zeiss; magnification x400).

*2.7. Statistical Analysis*

Analysis of variance (ANOVA) for each parameter quantified was performed to test the differences between experimental culture conditions (ihOEG cell monolayers and PLL). If differences were significant, post hoc Tukey test for multiple comparisons between means was carried out.

## 3. Results

*3.1. Infection and FACS selection of GFP positive Ts14.*

Ts14 and Ts12 lines derive from ihOEG hTL4 and hTL2 clonal lines, respectively, that were reversible immortalized with genes Bmi-1 and TERT (Fig. 1A shows Ts14 and Ts12 lineage). These lines express OEG typical markers: S100β (Fig. 1A), GFAP and Neuroligin [32,34].

Ts12 was obtained using SV40 large T antigen lentiviral transduction of hTL2 cell line, and selecting with the passages a population with rapid growing capacity (Fig 1A). Population doubling for these cells was determined to be around 20 hours.

Ts14 was infected with a lentivirus carrying E-GFP gene (Fig. 1A) and a positive population was isolated using flow cytometry and cell sorting (FACS). Ts14 cell line transduced with GFP represented 22.5% of the total population. FACS isolated Ts14-GFP cells were separated again using this technique to assure that all selected cells expressed GFP.

Both Ts12 and Ts14 presented nuclear expression of SV40 virus large T antigen (Fig. 1B and C).

*3.2. Axonal Regeneration Assay in vitro: coculture of RGN with ihOEG*

In one set of experiments, regenerative capacity of ihOEG line Ts12 was compared with the regenerative Ts14 line, in seven independent cocultures. RGN adherence/survival values were 6.38±0.67 for Ts14 and 9.79±1.98 for Ts12 (neurons/field, mean ± S.E.M.; Fig. 2A). Percentages of neurons with an axon were 15.41±1.40% and 6.20±0.96% for Ts14 and Ts12, respectively (mean ± S.E.M, $p < 0.001$, Anova and post-hoc Tukey test; Figs. 2B, 2D and 2E). Finally, axonal length/neuron in Ts14 and Ts12 were 47.08±5.82 and 12.88±4.11 µm/neuron, respectively (mean ± S.E.M, $p < 0.001$, Anova and post-hoc Tukey test; Figs. 2C, 2D and 2E). On PLL we obtained a high, but not significant (Anova) RGN adherence value (11.20±2.01 neurons/field, mean ± S.E.M, Fig. 2A), but very low axonal regeneration was observed, as assessed by percentage of neurons with an axon (2.86±0.33%; mean ± S.E.M., $p < 0.001$, Anova and post-hoc Tukey test, vs. value obtained for Ts14; $p = 0.07$ Tukey test vs. value obtained for Ts12; Figs. 2B and 2F) and axonal length/neuron (3.13±0.91 µm/neuron, mean ± S.E.M., $p < 0.001$, Anova and post-hoc Tukey test, vs. values obtained for Ts14; $p > 0.05$ Tukey test vs. value obtained for Ts12; Figs. 2C and 2F).

In another set of experiments, Ts14 and Ts14-GFP capacities to induce RGN axonal regeneration were compared in four independent cocultures, to determine if GFP gene insertion affected the neuroregenerative ability of this new line. RGN adherence/survival values were 8.89±0.81 for Ts14 and 8.86±1.21 for Ts14-GFP (neurons/field, mean ± S.E.M., $p > 0.05$, Anova and post-hoc Tukey test; Fig. 3A). Percentages of neurons with an axon were 13.04±1.32% and 12.44±0.67% for Ts14 and Ts14-GFP, respectively (mean ± S.E.M., $p > 0.05$, Anova and post-hoc Tukey test; Figs. 3B, 3D and 3E). Finally, axonal length/neuron on Ts14 and Ts14-GFP were 38.47±8.89 and 34.62±9.10, respectively (µm/neuron, mean ± S.E.M. $p > 0.05$, Anova and post-hoc Tukey test; Figs. 3C, 3D and 3E). Newly, in PLL we obtained a high RGN adherence value (16.91±3.09 neurons/field, mean ± S.E.M. $p < 0.05$, Anova and post-hoc Tukey test vs. values obtained for Ts14 and Ts14-GFP, Fig. 3A), but low axonal regeneration, as assessed by percentage of neurons with an axon (1.13±0.65%; mean ± S.E.M., $p < 0.001$, Anova and post-hoc Tukey test vs. values obtained for Ts14 and Ts14-GFP; Figs. 3B, 3F) and axonal length/neuron (1.70±1.02 µm/neuron, mean ± S.E.M., $p = 0.016$ and $p = 0.028$, Anova and post-hoc Tukey test vs. values obtained for Ts14 and Ts14-GFP, respectively; Figs. 3C, 3F).

## 4. Discussion

We have previously described ihOEG lines to promote adult RGN axonal regeneration *in vitro* [32-34]. These human glial lines enable us to study the molecular mechanisms responsible for their neuroregenerative properties: in the case of hTL4 line we identified the pathway of PAR-1 and, associated to this, the role of PAI-1 [27]. However, hTLs growing rate in culture is similar to the primary cultures, with a population doubling time of ten days, approximately [32]. For this reason we have modified some hTL lines by viral transduction of the SV40 Large T antigen. In this way, we originated from hTL4 and hTL2 the lines arbitrarily named Ts14 [34] and Ts12 (described here), respectively. Both lines have a high growing capacity, with a population doubling time around 20-24 hours. Therefore, Ts14 provides an excellent tool to study ihOEG regenerative properties, both *in vitro* [34] and in animal models of CNS lesion.

Here, we aimed to modify Ts14 line, engineering it to constitutively express GFP. With this modification we will be able to track these regenerative cells in transplantation models after spinal cord and optic nerve injury. Then, we have evaluated the regenerative capacity of Ts14-GFP in comparison with its parental line Ts14, and using Ts12 as a low regeneration control cell. We demonstrate that the insertion of the GFP gene in the Ts14 genome did not change the regenerative properties of the new line, and that this is also true when we compare it with our other previously described ihOEG [32-34]. Thus, this new Ts14-GFP line is adequate for testing its neuroregenerative capacity and for cell tracking in *in vivo* models of CNS injury. By contrast, we have demonstrated here that Ts12 is a line that presents low neuroregenerative ability in our model of coculture with RGN, as its parental line hTL2 [32].

Previously, our group found differences after OEG immortalization between hTL lines, being precisely hTL2 a low-regenerative line. OEG primary cultures are heterogeneous, and because hTL2 is a clonal line, we cannot discard the random selection of a non-regenerative cell of the original culture. However, other possible explanation is that insertion of "immortalizing" genes in the genome of the original OEG cell may have disturbed genetic information fundamental for hTL2 regenerative capacity [32]. The additional engineering of this line to express SV40 large T antigen did not modify the lack of neuroregenerative capacity, as expected. Because gene transfer mediated by lentivectors does not permit to control integration sites or copy number insertion, a proof of concept must be performed: newly modified selected lines must be evaluated to assess preservation of their neuroregenerative properties.

In conclusion, i) our work provides us with a low-regenerative ihOEG line as a control for axonal regeneration studies (*in vitro* and *in vivo*); and ii) this study constitutes a proof of concept and confirms that the ihOEG line Ts14-GFP maintains the ability to induce axonal regeneration, as its direct parental Ts14 line, and it will be selected for *in vivo* studies in animal models of CNS lesion and transplantation.

**Acknowledgements**

This Project has been sponsored by Mapfre Foundation and by Universidad Francisco de Vitoria. We thank to Mr. Adrián Sánchez for his contribution to determine population doubling times for Ts14 and Ts12.

**FIGURE LEGENDS**

**Figure 1**. Human Immortalized OEG (ihOEG) lines, named hTLs, were obtained by reversible immortalization with Bmi1 and TERT genes, from primary human OEG cell culture [32]. Afterwards, hTL4 [34] and hTL2 were modified by lentiviral transfer with SV40 large T antigen, originating in this way Ts14 and Ts12 ihOEG lines (A). Finally, we have additionally modified Ts14 line to constitutively express green fluorescent protein (EGFP: Enhanced Green Fluorescent Protein). S100β and GFP immunostaining, and DAPI nuclear staining of primary culture cells, are shown in the figure (A). ihOEG lines Ts12 (B) and Ts14 (C) in culture express nuclear SV40 virus large T antigen. Bars 50 µm.

**Figure 2.** Quantification of axonal regeneration was performed in cocultures of RGN with Ts14 and Ts12. Number of neurons per field (A), percentage of neurons with an axon (B), and mean axonal length, expressed as µm/neuron (C) were quantified (PLL: Poly-L-Lysine, 10 µg/ml). Means ± S.E.M. of 7 experiments, with duplicated samples for each experimental condition, are shown. Asterisks indicate the statistical significance: ***$p < 0.001$ (ANOVA and post hoc Tukey test comparisons between parameters quantified for Ts14 vs. Ts12 and for Ts14 vs. PLL. There was not statistical difference between the numbers of neurons per field in the different conditions). Cocultures of RGN with Ts14 (D) or Ts12 (E) and RGN on PLL 10 µg/mL (F) are shown. White arrows indicate RGN axons (SMI31-positive: green) and blue arrows neuronal bodies and dendrites (514 positive: red and yellow). Bar 50 µm.

**Figure 3.** Quantification of axonal regeneration was performed in cocultures of RGN with Ts14 and Ts14-GFP. Number of neurons per field (A), percentage of neurons with an axon (B), and mean axonal length, expressed as µm/neuron (C) were quantified (PLL: Poly-L-Lysine, 10 µg/ml). Means ± S.E.M. of 4 experiments, with duplicated samples for each experimental condition, are shown. Asterisks indicate the statistical significance: ***$p < 0.001$ and *$p < 0.05$ (ANOVA and post hoc Tukey test comparisons between parameters quantified for Ts14 and Ts14-GFP vs. PLL. There was not significant difference between values obtained for Ts14 and Ts14-GFP). Cocultures of RGN with Ts14 (D) or Ts14-GFP (E) and RGN on PLL 10 µg/mL (F) are shown. White arrows indicate

RGN axons (SMI31-positive: red) and blue arrows neuronal bodies and dendrites (514-positive: green). Bar 50 µm.

# FIGURE 1

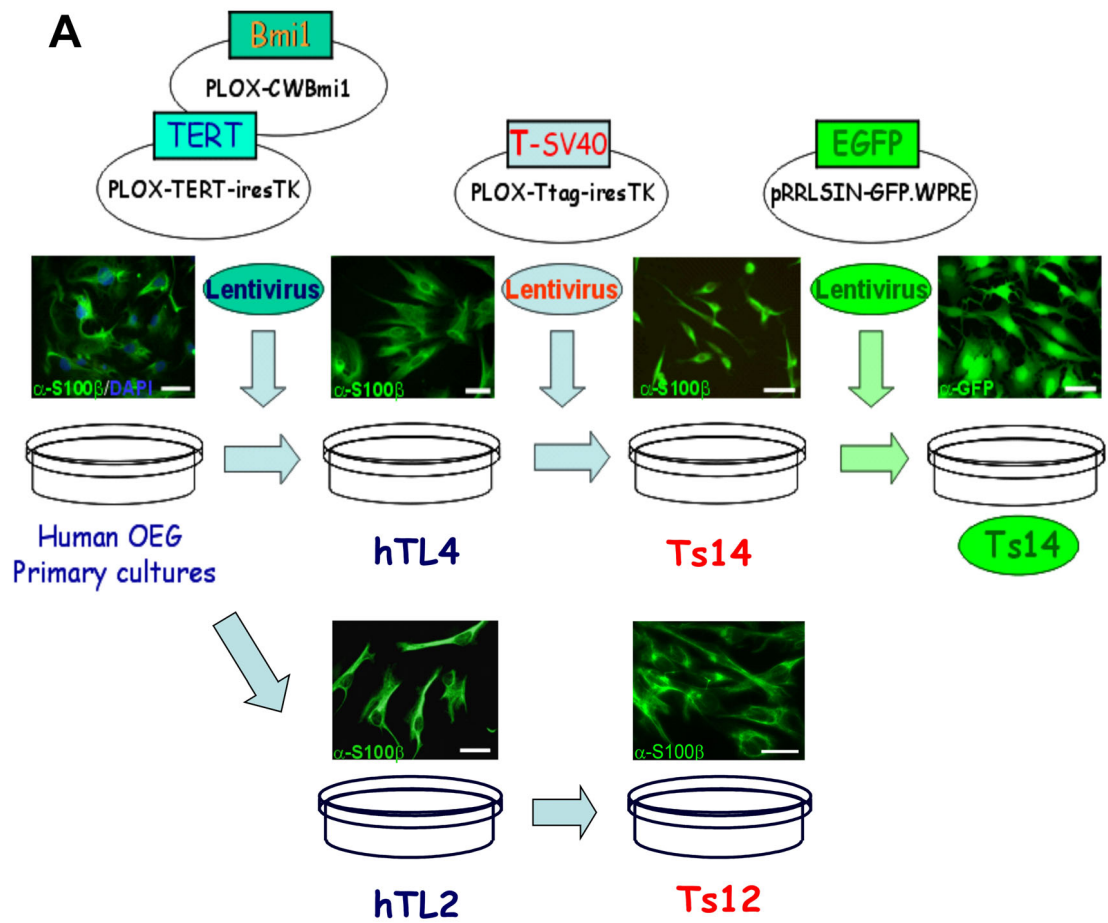
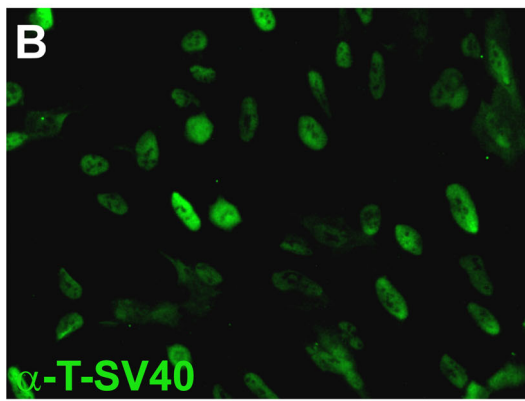
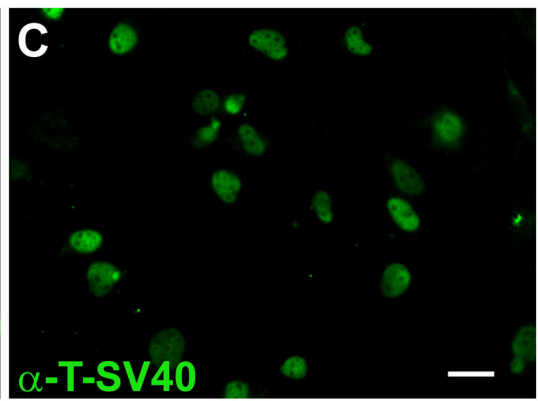

**FIGURE 2**

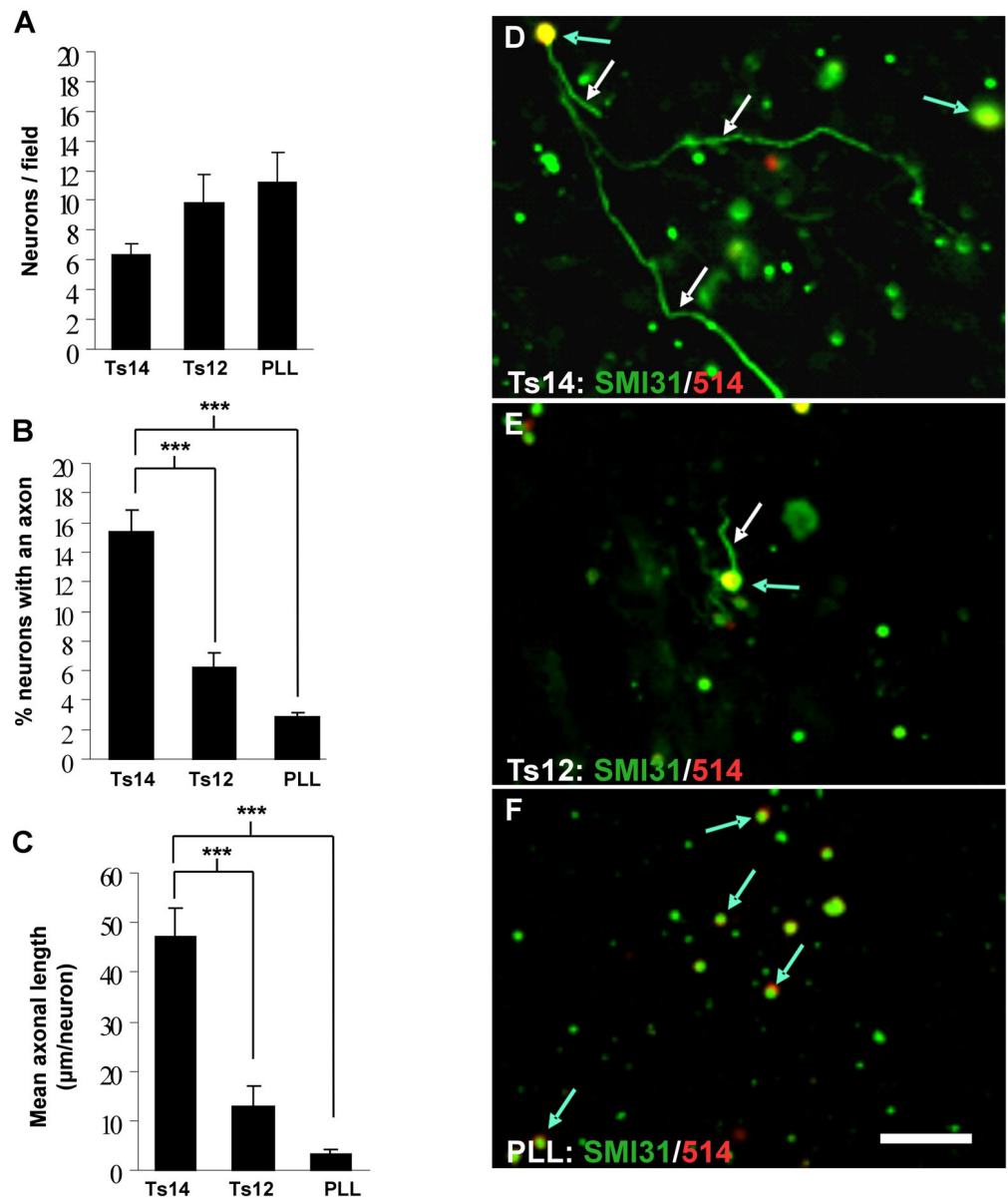

**FIGURE 3**

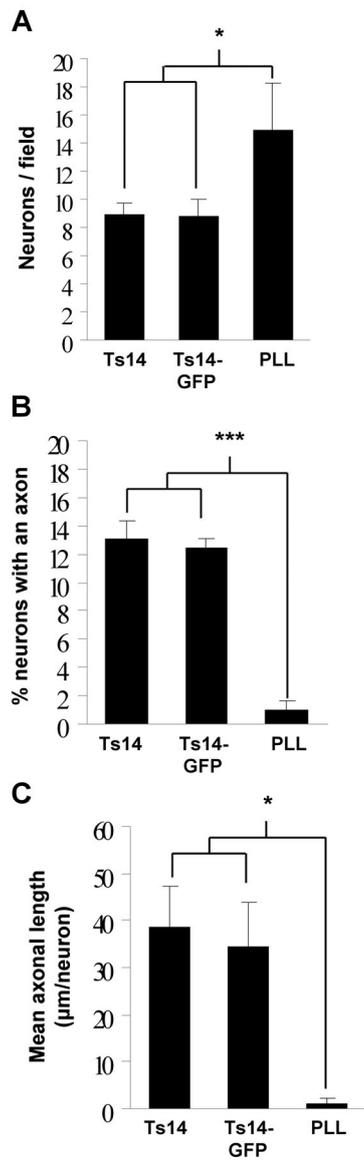
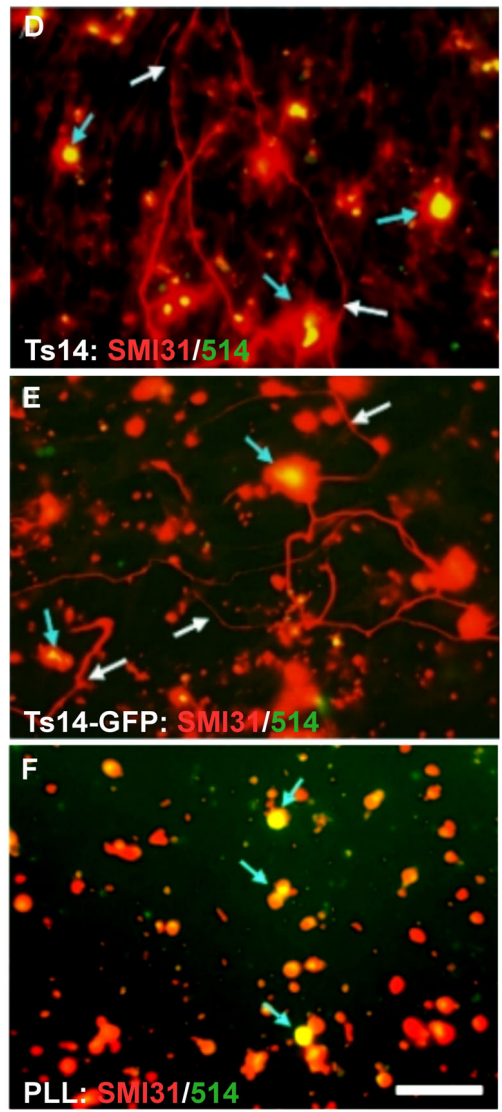